# A decentralized FAIR platform to facilitate data sharing in the life sciences

Pavel Vazquez[1*], Kayoko Shoji[1], Steffen Novik[2], Stefan Krauss[1] and Simon Rayner[3,1]

1. Hybrid Technology Hub Centre of Excellence, Institute of Basic Medical Sciences, University of Oslo, Oslo, Norway
2. Department of Informatics, Faculty of Mathematics and Natural Sciences, University of Oslo, Oslo, Norway
3. Department of Medical Genetics, Oslo University Hospital and University of Oslo, Oslo, Norway
*Corresponding author

**Dictionary**
GADDS: Global Accessible Distribution Data Sharing
Machine: physical hardware that can execute commands.
Node: machine in a network.
Cluster: group of machines.
Organization: group of nodes sharing a domain name.
Domain: network address.
Channel: permissioned network where organizations communicate.
Consortium: group of organizations that share a common channel.
(Meta)data: data and metadata.
Project: data and metadata duple

## Abstract

The Hybrid Technology Hub and many other research centers work in cross-functional teams whose workflow is not necessarily linear and where in many cases technology advances are done through parallel work. The lack of proper tools and platforms for a collaborative environment can create time lags in coordination and limited sharing of research findings. To solve this, we have developed a simple, user-friendly platform built for academic and scientific research collaboration. To ensure FAIRness compliance, the platform consists of a metadata quality control based on blockchain technologies. The data is stored separately in a distributed object storage that functions as a cloud. The platform also implements a version control system; it provides a history track of the project along with the possibility of reviewing the project's development. This platform aims to be a standardized tool within the Hybrid Technology Hub to ease collaboration, speed research workflow and improve research quality.

## Introduction.

In the last two decades, the life sciences have been revolutionized by technical advances in experimental methodology[1]. Nowadays, researchers not only generate huge amounts of data in a single experiment but the types of data they are collecting has also become highly divergent. There is also a need for data descriptors, i.e. metadata, to supplement the raw data that is collected as part of an experiment. Thus, researchers need to store metadata and data, or (meta)data as a standard step in an experiment. Hence, biology is making the transition towards a data science and a 'life cycle' view of research data[2]. Researchers now face the challenges associated with handling large amounts of heterogeneous data in a digital format. Some of these challenges include consolidating the data; translating it into a format that can be read by complex analysis pipelines; determining the most suitable analysis parameters; and making the data publicly available for reuse. There is growing evidence to suggest that many published results will not be reproducible over time [3]. Thus, robust data management and stewardship plans are essential to ensure the long-term sustainability of digital data.

The Findable, Accessible, Interoperable, Reusable (FAIR) data initiative was created in 2016 to address these issues by providing a framework for defining the minimum elements required for good data management[4].However, adopting FAIR principles is not straightforward as it requires knowledge of metadata, schemata, protocols, policies and community agreements. Moreover, the lack of exactness in the original FAIR principles means that there is an absence of clear implementation guidelines. Even when robust solutions exist, data providers may have to choose among different and not necessarily compatible implementations.
As publishers, funding agencies and policymakers are becoming increasingly aware of the FAIR data initiative, there have been efforts to implement measurable indicators of FAIRness[5].Nevertheless, for individual researchers, trying to incorporate FAIR data concepts in their data collection process remains challenging[3].

The Organ on a Chip research environment is recognised as a key emerging technology [6]. Organ on a Chip seeks to simulate the activities, mechanisms and physiological response of organs or organ systems. A major data challenge is that Organ on a Chip research collects huge amounts of highly diverse types of data that need to be integrated to understand the mechanics of an organoid design. However FAIR concepts have yet to be incorporated and currently no standards exist in the field. In addition to the challenges of integrating the data, there is also the problem of how to compare results among different research groups. For example, there are several Liver on Chip designs [7-9], but no way to compare performance.
In this paper, we introduce the Global Accessible Distribution Data Sharing (GADDS) platform (https://bitbucket.org/pavelva/gadds), an all-in-one cloud platform to facilitate data archiving and sharing with a level of FAIRness. The GADDS platform uses decentralization technologies and a tamper proof blockchain algorithm as a metadata quality control. By providing a browser-based client interface, the GADDS platform can simplify the implementation of FAIRness in the data collection and storage process. The platform is specifically developed for the Organ on Chip environment but has general application in any data collection and integration process requiring a level of data FAIRness. The GADDS platform integrates version control, cloud storage and data quality control as an all-in-one platform. In this paper, we present the motivation, conceptualization and the architecture of the GADDS platform and demonstrate how it facilitates data archiving and sharing in a FAIRlike Organ on a Chip environment.

## Results

**Architecture overview.**

The GADDS platform is intended to be deployed as a geographically distributed enterprise to aid the sharing of standardised data among laboratories implementing diverse technologies but working towards a common goal. The GADDS platform is designed to be a global federation (i.e. a group of computing resources sharing similar standards) where instances of resources form a unified global namespace. In this way, the GADDS platform is able to support an unlimited number of distributed participants.

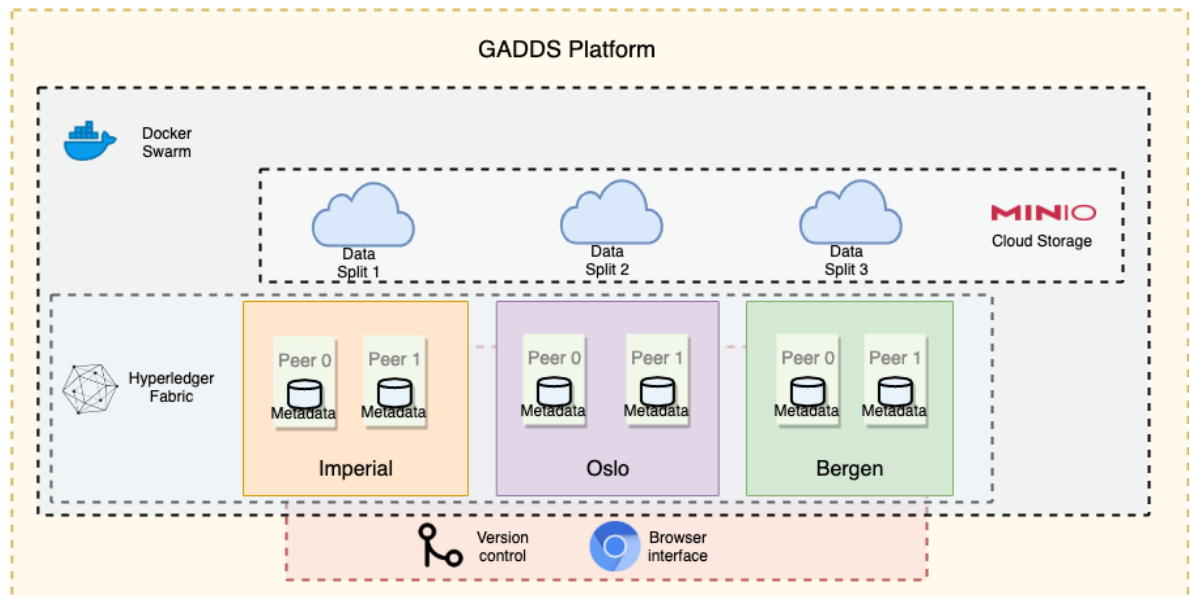

**Figure 1.** Schematic of the GADDS platform. The platform is deployed as a Docker Swarm cluster. The schematic shows a platform architecture distributed across three organizations (Imperial College Lonon, Oslo University Hospital and University of Bergen). Each organization hosts two peers that are responsible for metadata validation for quality control and to store the metadata.

The platform is based on three open source technologies:
1) Blockchain: provides a decentralized system to ensure that metadata standards are being followed.
2) Cloud storage: provides fault tolerant distributed storage, so that security and survivability of the data is improved.
3) Version control: allows the tracking of changes and recovery of different versions of (meta)data.

A schematic of the GADDS platform is shown in Figure 1. The whole platform is configured to be a Docker swarm cluster of a group of **machines** that can be either physical or virtual which execute commands in form of applications, and **nodes** are the machines that have been configured to be joined in a network. The activities of the cluster are controlled by a node called a Docker swarm manager but for simplicity, in the GADDS platform we have configured all nodes to be managers.

Figure 1 shows the configuration of the different environments that compromise the GADDS platform. The data storage of MinIO[10] and the Hyperledger Fabric[11] are part of the Docker[12] environment. In our implementation, we have configured three organizations in three different locations: Imperial College London, Oslo University Hospital, and nodes from the University of Oslo located in the city of Bergen. The version control and the browser interface are part of the Hyperledger environment but do not participate in the blockchain.

Following the Hyperledger architecture definitions, an **organization** is one or more nodes that share the same **domain name**, a **channel** is a *permissioned* network where different organizations communicate, and organizations that share a common channel are called **consortiums**. An advantage of having the flexibility of different architecture configurations is that organizations can share data in a secure manner and be geographically distinct, making a catastrophic event, such a server failure or an intrusion, to be an isolated event.

In Hyperledger, there are different nodes with different functions: the **peers**, that participate in the metadata validation and store metadata; the **certificate authorities** that are responsible for the permissions and one **ordering** node on each organization, that is in charge of the appending the metadata into the ledger. For simplicity we have chosen to eliminate these nodes in this implementation of the GADDS platform and instead we generate the permissions beforehand.

As the GADDS platform is a permissioned environment, only users from the same consortium can download data. In Figure 1 all organizations belong to a single consortium, thus all participants can share the data.

The flexibility of Docker allows the architecture and configuration of the GADDS platform to be changed, so peers and organizations can be subsequently be added or eliminated. Nevertheless, each organization maintains the same functionalities.

As shown in figure 1, data and metadata are stored separately within each organization. Peers participate in the blockchain and store the open metadata, while specific nodes in the organizations store the data in a secure environment. A more detailed discussion about the blockchain functionalities is presented in the *Methods* section.

**(Meta)data lifecycle**

The users act as peers and interact with the GADDS platform through a simple web interface where the (meta)data has been separated but linked together by a unique identifier, so a (meta)data duple is considered an **experiment**, where each experiment has a name and an identifier. A collection of related experiments forms a **project** which is placed inside a **bucket**.

Figure 2 shows a schematic of the data lifecycle. In the data upload step all three components (i.e. the blockchain, the cloud storage and the version control) of the GADDS platform participate. First the metadata needs to be validated by the blockchain consensus algorithm (see Methods section) that operates among peers within the consortium (left hand side Figure 2). Once the metadata is validated by consensus, the metadata is incorporated into a block and is appended in the open ledger, i.e. the ledger is composed of metadata entries in the form of blocks.

After the metadata has been validated by the blockchain, the data is uploaded as an object into a bucket in the cloud storage, at the same time a snapshot of the (meta)data with a timestamp is generated by the version control.

Metadata searching and data download can only be performed by peers within a consortium. When searching the ledger, the system will only return the metadata results for which the user has read permissions. Similarly, when the user attempts to download the data associated with selected metadata hits, a verification step is performed to ensure they have access permissions. Here, in the example GADDS configuration, the metadata and data permissions are identical.

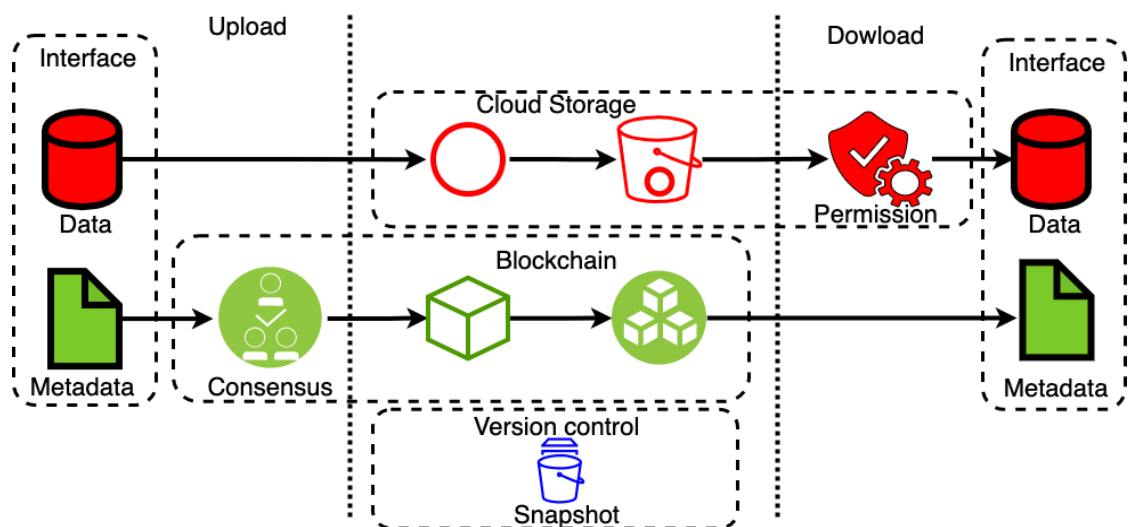

**Figure 2.** (Meta)data lifecycle

(Meta)data can also be modified in experiments, which leads to the creation of a new duple. When a request is made to modify an experiment a new (meta)data duple is created, in which the experiment name and identifier are maintained from the original entry. This new metadata will then be verified by consensus and, if successful, will be incorporated into a new block. Thus, the new metadata will point to the original data object. Each modification creates a snapshot in the Version control, so a history of all the modifications is kept in a history log, which is saved in the peer's personal hard drive and in the cloud storage.

**Interface and use case**

In order to prove that the GADDS platform can work effectively in a research environment itself, here we show a simple demonstration using a simplified Organ on Chip experiment consisting of a single dataset. This demonstration shows the potential for the GADDS platform to enhance the collaboration among different groups within the Hybrid Technology Hub.

First, the fabrication of microfibres was performed by the Tissue Engineering group, that produced the core-shell hydrogel fibres following the procedure described in [13]. The fibres were formed using a co-axial nozzle made of glass capillaries. ECM protein solution +/- cells and sodium alginate solution were introduced into the nozzle respectively at certain flow rates to form co-axial laminar flow in the nozzle. Then the flow was ejected in 100 mM calcium chloride solution to solidify alginate thus forming the core-shell. The data that will be stored in to the platform was: 1) composition of solution used as central layer (including which ECM proteins and cells, concentration of the protein and the cells) and 2) concentration and type of sodium alginate solution.

The diameter of the outer shell and inner core of the fibres could greatly impact the final outcome of future experiments. After the fibre fabrication, the group analysed the outcome by measuring both the outer and inner diameters. The group compiled an illustrative dataset containing the diameter of multiple fibres measured on three different locations along the fibre's axis. The dataset stored in the platform was: 1) inner and outer diameters of the glass capillaries of the nozzle used and 2) flow rates. This dataset has been used to produce valuable statistics of the variability of the core- and shell diameters.

Before submission of data the system requires a minimum metadata set of elements to be verified by the platform. We have chosen the Dublin core metadata[14] standard of 15 elements as a straightforward descriptive standard to follow our example, this standard is used as the minimal amount of information that allows datasets (or other resources) to be easily discovered and identified, providing an extended Interoperability for the uploaded dataset.

The 15 Dublin metadata core elements have been coded into the GADDS as a default template; when uploading the dataset, the platform verifies through the metadata quality control system that the relevant metadata elements are input in the proper standard and that entries for all the elements are present, e.g. the language element must follow the ISO 639-1 standard of two letters. This experiment shows the proposed potential of the GADDS platform in a real-world application showing that the platform can improve collaboration within a cross-disciplinary research environment.

## Methods

**Architecture in more detail.**

The GADDS platform is designed to be a fault tolerance system by using multiple distributed system architectures, which provides (meta)data survivability while at the same time oversees metadata quality assurance. The GADDS architecture has been designed keeping in mind that computing systems are bound to fail so it will continue to provide its specified function despite the occurrence of failures.

As we have mentioned, the GADDS is a hybrid of three technologies, blockchain for decentralization and fault-tolerance in the quality control system, cloud object storage for a distributed fault-tolerance to store the data, and a versioning system to track changes in data. In the following sections we will describe these three components in more detail.

**Docker**

We use Docker containers and swarm technology for its ease of use and flexibility, this is especially useful when deploying packaged code across multiple computers in a test environment, where code changes are needed to be done frequently. In Docker, a container acts as an independent package of software, this is useful as we can ensure that the platform can work uniformly despite differences among machines with different hardware or operative system. The Docker Swarm simplifies the management of the containers across multiple machines, it manages the resources to ensure they are handled efficiently.

**Blockchain and Hyperledger Fabric.**

Blockchain is the backbone of the cryptocurrency BitCoin[15]. As it was conceptualized and applied from its beginnings, the blockchain is a distributed database that records all transactions (exchanges) that have been executed among participants. Each one of these transactions is endorsed by the majority of participants of the system through an algorithm. This database, called the ledger, contains every single record of each transaction in the form of blocks.

One of the main characteristics of the blockchain is that its functionality is decentralized, meaning that there is no central system which keeps the records; thus, the ledger is distributed over all the participants which are connected with the blockchain. Thanks to its decentralization, blockchain enhances trust, the idea is not that participants in a transaction cannot trust those who they conduct business with, it's that they don't have to when operating on a blockchain network. Some relevant characteristics of the blockchain are:
- Decentralization: the open ledger is shared and updated with every node connected to the blockchain. Participants must agree that a transaction is valid, and this can be achieved through the use of consensus algorithms.
- Security: access to the blockchain is through permissions and cryptography.
- Transparency: every node in the blockchain has a copy of the ledger.

The blockchain operates using a consensus algorithm, this is a procedure where all the peers of the blockchain network reach a common agreement about the state of the ledger. In the GADDS platform, the consensus is an agreement about whether to include specific metadata into the ledger. Essentially, the consensus makes sure that every new block in the ledger that is added is the one and only version that is agreed upon by all the participants. In this way, the consensus algorithm establishes trust between peers in a distributed computing environment.

A common consensus algorithm is Proof of Work (PoW), the idea behind this algorithm is to solve a complex mathematical puzzle. This mathematical puzzle requires a lot of computational power. In cryptocurrencies there is a reward for the participants that help to solve the puzzle. These participants are called the miners. In the GADDS platform we use the Hyperledger Proof-of-Authority (PoA) consensus[16], this is different from PoW, here the right to participate and generate new blocks in the ledger is awarded to nodes that have proven their authority and identity. These nodes are the "Endorsers-Validators" and they run specific software that allows their participation. The value of these nodes is their identity, in contrast to PoW, the stake for these nodes is their "reputation". In order for the blockchain to maintain its integrity, validators need to confirm their real identities, thus they need to be verified by their peers.

Some advantages of using PoA is that transactions, in our case metadata validations, do not require computational power, so the rate of validations is much faster than in the PoW and less computationally expensive. PoA has also a high-risk tolerance, as long as 51% of the endorser-validators are available and not acting maliciously.

As we have mentioned, in the GADDS, the blockchains' function is metadata quality control while also acting as a database for metadata. In the GADDS context, the PoA consensus algorithm validates the input of metadata to ensure that is following a predefined standard (see data standards). If the endorsement and validation of the metadata is successful, it then becomes a block and it is appended to the ledger. This means that metadata is stored in the ledger in the form of blocks, each with a unique block ID (MID) which is different from the identifier that links the block with its corresponding data (DID), this is useful when linking the metadata with its corresponding metadata.

The GADDS platform follows the Hyperledger Fabric mechanism to validate the metadata. Here we will describe, at a conceptual level, how the algorithm allows organizations to collaborate in the formation and validation of metadata.

The main concept of the Hyperledger Fabric is that the metadata validation is being made by dedicated nodes, most notably the endorsers, committers and validators. In the GADDS platform, we have kept these three nodes jointly, while the so called Orderer has been kept separated.

The metadata validation mechanism starts when a user submits (meta)data through the interface and ends when the metadata has been stored in the ledger and the data stored in the cloud. The following steps occur during the metadata quality control:

Step 1.1: The web interface package sends a request to the EVC nodes to start the endorsement process.
Step 1.2: The web interface packages the metadata as a transaction and sends it to the peers.
Step 2.1: Each EVC node performs an endorsement of the credentials of the interface.
Step 2.2: At the same time, each EVC node performs an endorsement of the metadata by comparing against a template.
Step 2.3: The EVC nodes send their endorsement responses to the Orderer.
Step 2.4: This is repeated for several submitted metadata in form of transactions.
Step 3: The Orderer gathers several transactions and packages them into a block.
Step 4: The Orderer sends the assembled block to all EVC nodes.
Step 5: The EVC nodes validates each transaction within the block by consensus.
Step 6: Each transaction within the block is updated with the result of the validation.
Step 7: Each peer adds the block to its copy of the ledger.
(this means that a block may contain one or more invalid metadata, but these are filtered out when a user queries the ledger).

During step 2.2 a specific code is invoked; this code is called the chaincode and acts similar as a transaction in the blockchain. The chaincode sole function is to automatically compare the metadata with a predefined standard in the form of a template.

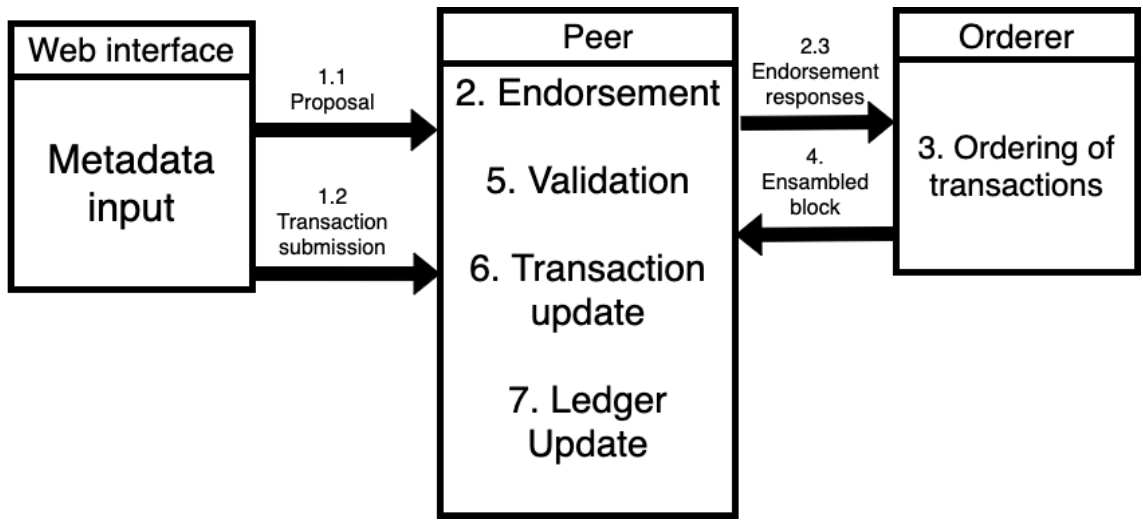

Figure 3. Metadata validation flow.

Upon the execution of a transaction, it is necessary to specify the peers and the channel that the transaction will be executed on. The use of channels is to provide a completely separate communication layer between participants, but channels also provide a way of sharing a network between multiple participants while maintaining data and communication privacy. As we have mentioned before, in the GADDS platform, we have defined the participants within a single channel in single consortium.

Hyperledger Fabric has the capability to have two specialised nodes called the Membership service provider (MSP) and the Certificate authority (CA) that can actively manage identities and issue the corresponding certificates in order to grant permission to participants, as for the moment, we have predefined upon start the identities throughout certificates.

**Cloud storage**

Rather than dedicated servers used in traditional networked data storage, in the GADDS platform we use a cloud architecture where data is split, replicated and stored on multiple servers. This technology is similar to Dropbox or Google Drive where the data is redundantly stored, often across multiple facilities, so that natural disasters, human error, or mechanical faults do not result in data loss. Thanks to cloud storage technologies, even if the data is split the users see what corresponds to a single file on the server as if the data is stored in that particular place within a specific namespace.

The GADDS platform implements MinIO as cloud storage technology. We have opted for this solution as it is open source, relatively straight forward to deploy through a Docker swarm and well documented. MinIO splits and replicates the data into "chunks" or parts, see Figure 4, this helps to protect the data against failures such as corruption, hardware failure or intrusions by using "Erasure code"[17]. Thanks to this a high level of redundancy is achieved, it is possible to lose up to half (N/2) of the total storage devices and still be able to recover the data[18]. However, high redundancy also means higher storage usage.

This specific cloud storage solution follows object storage architecture, that manipulates data as units called objects. From an end user perspective an object corresponds to a file. Object storage combines pieces of data that make up a file and adds file specific metadata to that file, while attaching a global and unique identifier. The GADDS platform uses this unique identifier to link the data submitted with the metadata in the blockchain.

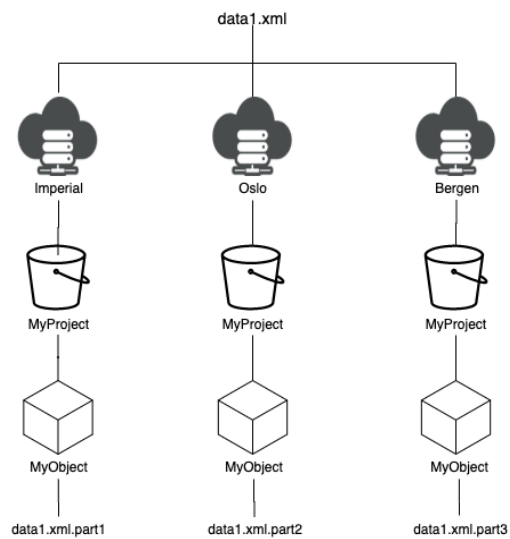

Figure 4. Four-organization example of the cloud object storage solution in a. A file is divided and replicated among organizations.

This scheme has the advantage that each part of the file is distributed in different secure environments, so each organization is responsible for the security and longevity of their nodes. We have used MinIO not only as a strategy to support disaster recovery, but also as a participation scheme where organizations share storage resources and ensure secure environments.

**Version control**

As part of the distributed design of the GADDS platform, the version control is developed to work with the MinIO storage solution to handle (meta)data submits and changes/updates. In this way, projects can scale up to large numbers of participants for geographically-distributed research. This system is capable of recording changes made to (meta)data entries, thus is possible to make modifications to submission that have been already validated. When submitting a change to the (meta)data through the web interface, the process of validation is initiated as if it was a new entry, if successful a new block of metadata is created. The older version of the block and the corresponding data are kept in the system, so just like Git[19], it is possible to go back in time to recall a specific version of that entry.

When first submitting and validating the (meta)data, the system creates a unique data identifier (DID), different from the metadata identifier (MID), that links the metadata (in the form of a block) with the data, see Figure 5. If there is a request to change either the metadata or data, the block identifier changes but not the unique identifier that links the data.

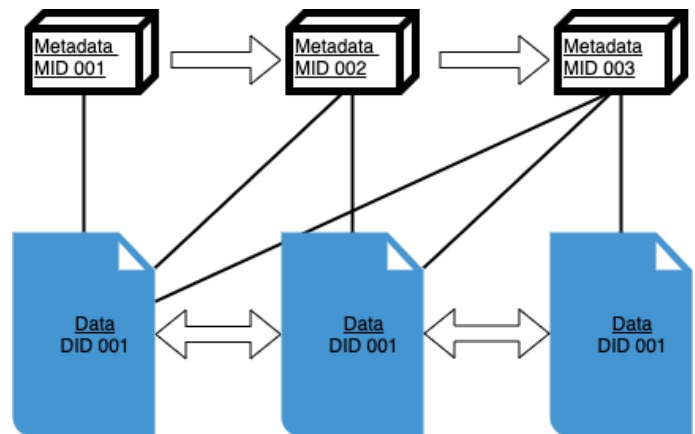

Figure 5. Version control. The metadata blocks are linked to the corresponding data.

The version control gives access to historical versions of the data. This is an insurance against computer crashes or data loss. If there is a mistake, it is possible to roll back to a previous version. Similar to Git, each user is able to edit his or her own copy of the data and chooses when to share by submitting both metadata and data. Thus, temporary or partial edits by one user is not visible in the GADDS until the meta(data) is submitted. Once the (meta)data is submitted it is possible for one person to use multiple computers to edit that entry. At the current state of the version control, if two users edit the same metadata the latest submission will be the one available to retrieve.

## Discussion

The GADDS has been strongly inspired by the reusability of data and the implementation of FAIR principles to scientific datasets while incorporating the philosophy of the cloud storage and decentralization technologies. Thus, the platform combines the inherent advantages of the blockchain and the object storage in a suite of features with a simple interface.

We have aimed the GADDS platform towards easing collaboration and data sharing while being inspired on the FAIR initiative. This initiative is a guideline to help data to be Findable, Accessible, Interoperable and Reusable. It was first described in 2016 [4] with the intention to provide a framework that defines basic elements required for good data management. The principle of **Findability** stipulates that data should be identified, described and registered or indexed in a clear and unequivocal manner; in the GADDS we have made a requisite that data is described with relevant metadata while using unique identifiers. The principle of **Accessibility** proposes that datasets should be accessible through a clearly defined access procedure, ideally by automated means; in the GADDS platform the metadata is open and free, while the data follows a process of authentication and authorization. The principle of **Interoperability** mentions that data and metadata are conceptualised, expressed and structured using common standards; the metadata in the GADDS uses common standards and vocabularies. Finally, the principle of **Reusability** indicates that the characteristics of the data should be described in detail according to domain-relevant community standards, including clear and accessible conditions for use; we have procured that the metadata in the GADDS has the relevant attributes while meeting relevant standards, while metadata and data and linked together with unique identifiers.

The FAIR initiative is then a guideline to help, among other things, reproducibility of data by having well documented metadata and metadata standards. In the platform we have proposed the use of pioneering technologies to form a hybrid cloud platform. We have separately stored, through different architectures, the data from the metadata and linked them through unique identifiers. The data is stored using a distributed technology in the form of a cloud storage and in order to store metadata we use a consensus algorithm based on blockchain. The blockchain works as a metadata quality control in order to guarantee that researchers and research groups data comply a predefined metadata standard. In the current state of the GADDS platform, we have chosen built-in metadata standards but, as following feature, we will implement the possibility for the user to implement their own standard.

The adoption and implementation of the FAIR principles has proven to be a complex task that involves not only knowledge but also awareness of metadata, policies, and community agreements and other elements. The GADDS platform proposes a way to improve the process of better data management mainly by having open and well documented metadata and ensuring that predefined metadata standards are being followed. As for the moment, the platform gives restricted access to the data, and it is only accessible to users within the defined organizations. We propose a future development where the researcher can choose appropriate licencing characteristics to determine the openness level of the data to significantly increase the reusability and interoperability.

A we have mentioned, the GADDS is an all-in-one platform that assembles different pioneering technologies: a blockchain to store metadata, a cloud to store data and a version control system. The blockchain decentralization algorithm that we use is a novel initiative that encourages the use of metadata standards. We use blockchain technology as an open database in the form of a distributed ledger. The GADDS employs a permissioned blockchain by Hyperledger that implements a consortium of nodes tasked with creating new blocks made of metadata, while it executes a traditional Byzantine consensus protocol in order to decide which of the metadata, in form of blocks, are suitable to be inserted to the ledger (metadata database). The blocks are validated by consensus, hence the blockchain used by GADDS does not spend the amount of resources of other blockchains and is able to reach better transaction throughput.

While the metadata is stored as blocks in a decentralized ledger, the data is stored in a cloud system, this means that the physical storage is distributed among multiple servers in multiple locations. We have based the cloud on MinIO which is Amazon S3 cloud storage compatible, and that manages data as objects. The data is split and replicated among the different servers, so this technology allows the GADDS to avoid data loses in case of unexpected events such as power outages or hard drive failures but it also improves data security as each server can be localized in different security environments, thus in a case of a server breach, only partial data can be retrieved.

To coordinate work and facilitate collaboration among researchers we have implemented a distributed track changing system, i.e. a version control, with basic functionalities but similar to Git. The main purpose of this system is to keep a record of the changes that the data has experienced, so researchers can roll back and inspect data changes if necessary. Even though data collected might not be of non-linear workflows, researches can track their modifications of data if necessary.

There's the risk that the usage of restricted data can lead to the creation of data silos that cannot be used outside the consortium, but simply opening up the data to everyone does not suddenly turn the data into usable, so governance measures and policies need to be done/taken. The GADDS needs to take further steps in implementing a way to give accessibility to data outside the consortium. In that aspect, we also recommend that a proper body dedicated to data governance, in order to oversee the data management, should be taken in place.

The GADDS can help to simplify the process for data management but researchers need to make sense of their own data, so data literacy is also needed, maybe offering regular training to users is needed.

At the current version of the GADDS, in the platform's architecture a single Docker swarm cluster is used, the possibility to scale to multiple clusters in order to form a Federation is kept open. There are multiple benefits to a GADDS federation:
- There is a democratization of resources, each cluster has the same functionalities as the others.
- Metadata is shared among all nodes. Data provenance is open in the federation.
- The performance of an individual cluster remains constant as you add more clusters.
- Failures are kept in isolated parts. An issue with one cluster should not affect the entire Federation.

One "piece" of metadata corresponds to one file of data, at the same time metadata needs to be put manually but future developments for an automatization is proposed.